\documentclass[12pt]{article}
\usepackage[utf8]{inputenc}

\title{An impure solution to the problem of matching fans}
\author{Anton Salikhmetov}

\usepackage[utf8]{inputenc}
\usepackage[english]{babel}
\usepackage[title]{appendix}
\usepackage{fullpage}
\usepackage{csquotes}
\usepackage[
    backend=bibtex,
    style=numeric,
    block=ragged,
    url=false,
    sorting=none,
    firstinits
]{biblatex}
\usepackage{amsmath}
\usepackage{amssymb}
\usepackage{amsthm}
\usepackage{fixmath}
\usepackage{wrapfig}
\usepackage{caption}
\usepackage{subcaption}
\usepackage{tikz}
\usepackage{tikz-inet}

\tikzset{every node/.style = {node distance=0em, scale=0.8}}

\newcommand\fan[1]{\delta_{#1}}
\newcommand\hole{{[\phantom M]}}
\newcommand\fv{\text{FV}}

\newtheorem*{conjecture}{Conjecture}

\begin{document}
\maketitle

\begin{abstract}
We propose an algorithm to solve the problem of matching fans in interaction net implementations of optimal reduction for the pure untyped lambda calculus without use of any additional agent types.
The algorithm relies upon a specific interaction nets reduction strategy and involves side effects in one of interaction rules.
\end{abstract}

\section{Problem}

Matching fans is the main problem of implementation of optimal reduction in interaction nets~\cite{optimal}.
Existing solutions use so-called oracle which is implemented using bracket and croissant agents in BOHM~\cite{optimal} and delimiter agents in Lambdascope~\cite{lambdascope}.

Aiming to eliminate the overhead due to the oracle, we decided to limit our signature to the basic types only (abstraction, application, erase, and fan) and try to achieve the same behavior of fans as if the oracle were still present.

This paper continues our previous work towards optimal reduction without oracle~\cite{orwoo}.
Specifically, 1) we use the idea of identities attached to fans from~\cite{orwoo}, 2) restrict the interaction nets reduction strategy to needed reduction~\cite{neededred}, and also 3) allow side effects in one of interaction rules.
The main idea is to memorize identities of two different fans at the moment of their first interaction.
It has led us to an impure solution we present in this paper.

The presented solution has been implemented in software and is available in the MLC package at \url{https://www.npmjs.com/package/@alexo/lambda} (version 0.5.0) where it exists as the \texttt{abstract} encoding which is the default algorithm.

\section{Solution}

We work in the version of interaction calculus described in~\cite{neededred}.

The signature of our interaction system is
$$
\Sigma = \{p,\ @,\ \lambda,\ \varepsilon\} \cup \{\fan i\ |\ i \in \mathbb{N}\} \cup \{a_M\ |\ M \in \Lambda\} \cup \{r_{C\hole}\ |\ C\hole \text{ is a context}\}.
$$
Intuitively, one can think of $i$ in $\fan i$ as this agent's identity.
Agents $p$, $a_M$, and $r_{C\hole}$ are part of the embedded read-back mechanism described in~\cite[Section 7]{termgraph}.

The interaction rules are as follows:
\begin{align*}
r_{C\hole}[x] &\bowtie \lambda[a_y,\ r_{C[\lambda y.\hole]}(x)], \quad \text{where $y$ is fresh}; \\
@[x,\ y] &\bowtie \lambda[x,\ y]; \\
@[r_{M\ \hole}(x),\ x] &\bowtie a_M; \\
r_{C\hole}[a_{C[M]}] &\bowtie a_M; \\
r_{C\hole}[\fan i(x,\ y)] &\bowtie \fan i[r_{C\hole}(x),\ r_{C\hole}(y)]; \\
\fan i[a_M,\ a_M] &\bowtie a_M; \\
\fan i[\lambda(x,\ y),\ \lambda(v,\ w)] &\bowtie \lambda[\fan i(x,\ v),\ \fan i(y,\ w)]; \\
\fan i[@(x,\ y),\ @(v,\ w)] &\bowtie @[\fan i(x,\ v),\ \fan i(y,\ w)]; \\
\fan i[x,\ y] &\bowtie \fan i[x,\ y].
\end{align*}
Here, we omit the $\fan i \bowtie \fan j\ (i \neq j)$ interaction rule and will define it later explicitly extending reduction relation as it is this interaction rule that involves side effects.

We define a \textit{state} as
$$
c / (\varphi,\ n),
$$
where $c$ is a configuration as in~\cite{neededred}, $\varphi \subset \mathbb{N}^3$, and $n \in \mathbb{N}$.

The reduction relation on states is defined in two steps:
\begin{enumerate}
\item If $c_1 \rightarrow c_2$ as in~\cite{neededred}, then
$$
c_1 / (\varphi,\ n) \rightarrow c_2 / (\varphi,\ n).
$$
\item If $i \neq j$, then
\begin{align*}
\langle &!\fan i(t_1,\ t_2) = \fan j(u_1,\ u_2),\ \Delta \rangle / (\varphi,\ n)
\rightarrow \\
\langle
&t_1 = \fan {\varphi'(j,\ i)}(x,\ y),\
t_2 = \fan j(v,\ w),\\
&u_1 = \fan {\varphi(i,\ j)}(x,\ v),\
u_2 = \fan i(y,\ w),\
\Delta \rangle / (\varphi',\ n'),
\end{align*}
where $\varphi' = \varphi$ and $n' = n$ if $\exists x : (i,\ j,\ x) \in \varphi$, or
\begin{align*}
\varphi' &= \varphi \cup \{(i,\ j,\ n'),\ (j,\ i,\ j)\} \quad \text{and} \\
n' &= n + 1
\end{align*}
if $\nexists x : (i,\ j,\ x) \in \varphi$.
\end{enumerate}

(In MLC, $\varphi$ is implemented using a hash table.
As long as the latter's search and insert operations are $O(1)$, the cost of the $\fan i \bowtie \fan j\ (i \neq j)$ reduction remains $O(1)$, thus keeping number of reductions an adequate measure of efficiency.)

In order to encode a $\lambda$-term $M$ into a state, we first need to distinguish free variables from bound variables in $M$.
We do so by marking all free variables using the following operation:
$M^\bullet = M[\vec{x} := \vec{x^\bullet}]$, where $(\vec x) = \fv(M)$.

Then the following state is the initial encoding of a $\lambda$-term $M$:
$$
[M] = \langle r_\hole(!p) = x,\ \Gamma_n(M^\bullet,\ x)\rangle / (\varnothing,\ n),
$$
where $\Gamma_n(M, x)$ is the result of enumerating all $\delta$ occurrences in $\Gamma(M,\ x)$ as ${\fan 1,\dots, \fan n}$, and
\begin{align*}
\Gamma(x^\bullet,\ y) &= \{a_x = y\}; \\
\Gamma(x,\ y) &= \{x = y\}; \\
\Gamma(\lambda x.M,\ y) &= \{y = \lambda(\varepsilon,\ z)\} \cup \Gamma(M,\ z), \quad \text{when $x \not\in \fv(M)$}; \\
\Gamma(\lambda x.M,\ y) &= \{y = \lambda(x,\ z)\} \cup \Gamma(M,\ z), \quad \text{when $x \in \fv(M)$}; \\
\Gamma(M\ N,\ x) &= \{y = @(x,\ z)\} \cup \Gamma(M[\vec t := \vec {t'}],\ y) \cup \Gamma(N[\vec t := \vec {t''}],\ z) \cup \Psi(\vec t), \quad \text{where} \\
\Psi(\vec t) &= \{t_i = \delta(t''_i,\ t'_i)\ |\ t_i \in (\vec t)\} \quad \text{and} \quad (\vec t) = \fv(M) \cap \fv(N).
\end{align*}

The algorithm is correct if the following conjecture holds true for any $\lambda$-term $M \in \Lambda$.

\begin{conjecture}
$[M] \rightarrow^* \langle !p = a_N,\ \Delta\rangle / (\varphi,\ n)$ iff $N$ is the normal form of $M$.
\end{conjecture}

\section{Conclusion}

Our impure solution to the problem of matching fans makes use of the mechanism to track identities of fans suggested in~\cite{orwoo}.
We get rid of the na\"ive level-tracking part and utilize needed reduction~\cite{neededred}, blocking previously found counterexamples at the cost of side effects.

We believe that further work should be directed towards a pure version of the presented solution in order to investigate its properties and correctness.

\printbibliography
\end{document}